\documentclass[aps,pre,epsfigure,twocolumn,superscriptaddress]{revtex4-1}

\usepackage{graphicx}
\usepackage{amsmath}
\usepackage{latexsym}
\usepackage{amsfonts}
\usepackage{amssymb}
\usepackage{array}
\usepackage{epsfig}
\usepackage{color}
\usepackage[colorlinks=true,linkcolor=blue,urlcolor=blue,citecolor=blue]{hyperref}
\usepackage{hyperref}

\newcommand{\rhom}{\hat{\rho}}
\newcommand{\lan}{\langle}
\newcommand{\ran}{\rangle}
\newcommand{\Hop}{\hat{H}}
\newcommand{\nop}{\hat{n}}
\newcommand{\aop}{\hat{a}}
\newcommand{\adop}{\hat{a}^{\dagger}}
\newcommand{\Dop}{\mathcal{D}}
\newcommand{\im}{{\rm i}}

\newcommand{\tr}{{\rm tr}}
\newcommand{\lp}{\left(}
\newcommand{\rp}{\right)}

\newcommand{\la}{\lambda}
\newcommand{\La}{\Lambda}
\newcommand{\hr}{\lambda^+}
\newcommand{\Me}{M_t} 
\newcommand{\rd}{\rho^D} 
\newcommand{\Pm}{{{P}}}

\newcommand{\sutd}{Engineering Product Development Pillar, Singapore University of Technology and Design, 8 Somapah Road, 487372 Singapore}

\newcommand{\majulab}{MajuLab, CNRS-UNS-NUS-NTU International Joint Research Unit, UMI 3654, Singapore}

\newcommand{\ntu}{Division of Physics and Applied Physics, School of Physical and Mathematical Sciences, Nanyang Technological University, 21 Nanyang Link, 637371 Singapore}

\newcommand{\augsburg}{Institut f\"ur Physik, Universit\"at Augsburg, Universit\"atsstra\ss e 1, D-86135 Augsburg, Germany}

\newcommand{\nano}{Nanosystems Initiative Munich, Schellingstr. 4, D-80799 M\"unchen, Germany}

\newcommand{\NUS} {Department of Physics, National University of Singapore, 117542 Singapore, Republic of Singapore}

\begin{document}

\title{Occurrence of discontinuities in the performance of finite-time quantum Otto cycles}

\author{Yuanjian Zheng}

\affiliation{\sutd}

\affiliation{\ntu}

\author{Peter H\"anggi}

\affiliation{\augsburg}

\affiliation{\nano}

\affiliation{\NUS}

\author{Dario Poletti}

\affiliation{\sutd}

\affiliation{\majulab}

\begin{abstract}
We study a quantum Otto cycle in which the strokes are performed in finite time. The cycle involves energy measurements at the end of each stroke to allow for the respective determination of work. We then optimize for the work and efficiency of the cycle by varying the time spent in the different strokes and find that the optimal value of the ratio of time spent on each stroke goes through sudden changes as the parameters of this cycle vary continuously. The position of these discontinuities depends on the optimized quantity under consideration such as the net work output or the efficiency. 
\end{abstract}

\maketitle

\section{Introduction}
Recent years have witnessed a rapid growth in the study of heat engines operating at the nanoscale.  More generally, this area of research is increasingly progressing towards a multitude of energy efficient nano-technologies \cite{gemmerlnp2009,hanggirmp2009,sekimotolnp2010,gelbwasser2015,pekolanaturephysics2015}. On the experimental front, there have been several realizations of mesoscopic heat engines that employ a wide range of working fluids, albeit  operating almost exclusively in the classical domain. Typical examples include piezoelectric materials \cite{SteenekenVanBeek2011}, colloidal systems \cite{BlickeBechinger2012} and even a single atom \cite{RossnagelSinger2015}, to name but a few.

On the other hand, our theoretical understanding of quantum thermodynamics has undergone considerable development that has enhanced our ability to manipulate and control thermal devices at the nanoscale. For example, with the use of tailored driving protocols in various strategies collectively termed shortcuts to adiabaticity,  we are equipped to generate adiabatic or adiabatic-like dynamics in systems that are driven within a finite amount of time \cite{DemirplackRice2008, Berry2009, DengGong2013, PalmeroMuga2013, TorronteguiMuga2013, Jarzynski2013, DeffnerDelCampo2014, DelCampoPaternostro2014, DelCampo2013, AcconciaDeffner2015, DelCampoZurek2012, CampbellFazio2014, RohringerTrupke2015}. However their practical usefulness is  conditional upon the relative timescales of the cycle and the nature of the driving fields \cite{ZhengPoletti2015a}. More recently, studies of many-body working fluids in thermodynamic cycles are providing guiding principles that can enhance the performance of quantum heat engines. For instance, quantum statistics has been shown to significantly influence the work distribution of Hamiltonian processes \cite{YiTalkner2012, GongQuan2014}. In particular, the interplay between quantum statistics and other properties of the working fluid such as the trap geometry \cite{ZhengPoletti2015b} and/or many-body interactions \cite{JaramilloDelCampo2015} can result in augmenting the performance of a heat engine operating in the quantum regime. See \cite{gelbwasser2015,Kosloff2013, VinjanampathyAnders2015, KosloffLevy2014, GooldSkrzypczyk2015} for recent reviews on quantum thermodynamics and heat engines.

Furthermore, our understanding of heat engines at the nanoscale  has benefited from salient advances in statistical physics, namely the area of  fluctuation theorems \cite{hanggiNP2015,CampisiTalkner2011,CampisiTalkner2011err,Jarzynski1997,Crooks1999}. Among the various fundamental relations, we mention in particular the Jarzynski equality that has been validated experimentally in the classical regime, e.g. see Refs.~\cite{hanggiNP2015,CampisiTalkner2011,douarcheepl2005}. Although presenting a formidable challenge for experiment, the Jarzynski equality has also been verified also in the quantum regime \cite{batalhaoprl2014,AnKim2015}. These fluctuation relations allow us to explore various peculiarities in the thermodynamic behavior of non-equilibrium heat engines \cite{CampisiFazio2015a, CampisiFazio2015b}. Such peculiarities may arise from (but are not limited to) squeezed or non-thermal baths \cite{AbahLutz2012, AbahLutz2014, AlickiGelbwaserKilmovsky2015, Leggioantezza2016, MehtaPolkovnikov2013}, irreversibility \cite{JiangSegal2015,RezekKosloff2006}, finite-time effects of the driving \cite{GevaKosloff1992, FeldmannKosloff2000}, and more recently from the time-asymmetry used in the driving protocol \cite{GingrichGeissler2014,PalJayannavar2016}. Moreover, cycles that contain sudden changes in the Hamiltonian parameters have also been investigated \cite{FeldmannKosloff2003, FeldmannKosloff2012, FeldmannKosloff2015, UzdinKosloff2014}, thereby accounting for the role of noise in the cycles' performance as well \cite{FeldmannKosloff2006, AlecceZambrini2015}.

\begin{figure}
\includegraphics[width=0.9\columnwidth]{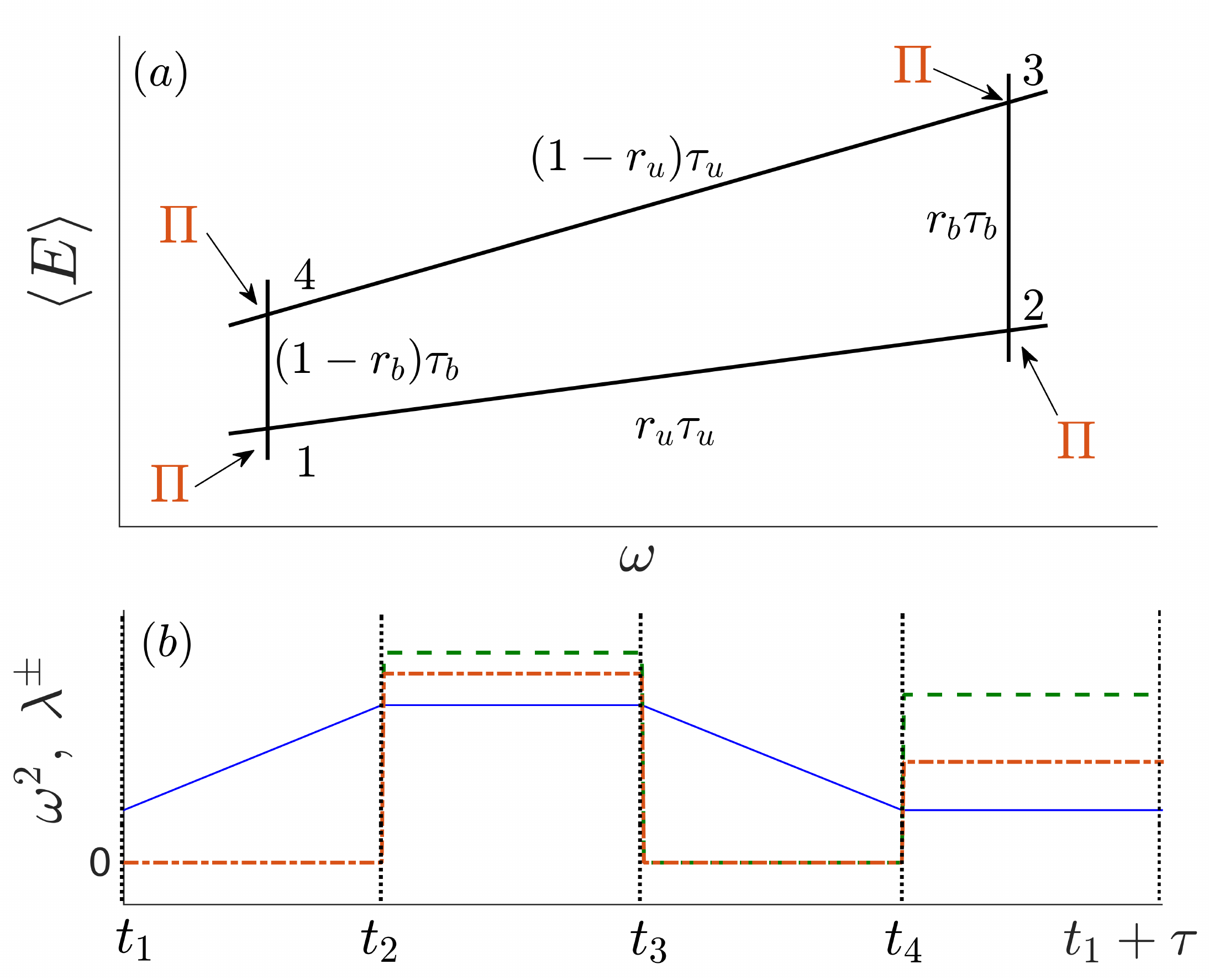}
\caption{(color online) (a)  {{\bf Schematics of a quantum Otto Cycle of total time duration} \boldmath${\tau=\tau_u+\tau_b}$}. Mean energy $ \langle E\rangle$ versus a time-dependent manipulation of the trap frequency  $\omega(t)$: $1 \to 2$. Compression: Unitary (subscript $u$) time evolution using a linearly increasing $\omega^2(t)$ during  the  time span $r_u \tau_u$. $2 \to 3$: Iso-parametric coupling to the hot bath (subscript $b$) at effective temperature $1/\beta_3$ during time $r_b\tau_b$. $3 \to 4$: Expansion: Unitary time evolution with linearly decreasing $\omega^2(t)$ during time interval $(1-r_u)\tau_u$. $4\to 1$: Iso-parametric coupling to the cold bath at effective temperature $1/\beta_1$ in remaining time span $(1-r_b)\tau_b$. Projective energy measurements $\Pi\lp\rhom(t_i)\rp $ are executed at the end of each stroke. (b) {\bf Time Dependence of Cycle Parameters.} The square of the angular frequency variation $\omega^2(t)$, depicted by the blue continuous line, increases linearly between $t_{1}$ and $t_{2}=t_{1}+r_u\tau_u$ while the bath coupling parameters $\lambda^{\pm}$, given by the red dot-dashed line ($\lambda^+$) and the green dashed line ($\lambda^-$), respectively, are held at vanishing values (zero system-bath coupling). Subsequently, after an instantaneous energy measurement and until time $t_{3}=t_{2}+r_b\tau_b$, both $\lambda^{\pm}$ are abruptly adjusted to  values which yield the hot effective temperature, $1/\beta_3$, while $\omega^2(t)$ is held fixed. After yet another energy measurement and till time $t_{4}=t_{3}+(1-r_u)\tau_u$, the angular frequency $\omega^2(t)$ is linearly reduced while the bath-coupling strengths $\lambda^{\pm}$ are again switched to zero. In the last stroke between $t_{4}$ and $t_{1}+\tau$ the angular frequency is held fixed while the $\lambda^{\pm}$ are turned to their new respective values corresponding to a lower effective temperature, $1/\beta_1$. Projective energy measurements are represented by vertical black dotted lines. Note that the ratio $\lambda^+/\lambda^-$ is different in the strokes indicating different effective temperatures in the two baths.}

\label{fig:1}
\end{figure}
\noindent

In this study, we further develop the understanding of non-equilibrium heat engines that operate during an overall time span $\tau$ upon employing strokes that are individually performed in finite time. In our analysis of the cycle, we introduce an explicit projective energy measurement which is performed before and after each stroke of the cycle, so as to determine the work via the two-time energy measurements protocol \cite{TalknerHanggi2007}. We then explore the non-equilibrium behavior by exploring the conditions for optimal work output and efficiency of the cycle and find characteristic discontinuities as a function of the system parameters.

In section (\ref{sec:model}) we describe the Otto cycle setup and detail the relevant parameters and figure of merits of the quantum engine. The Otto cycle is composed of four strokes: two unitary strokes, of total time $\tau_u$, intercalated by two strokes in which the system is weakly coupled to baths, for a total time $\tau_b$. Hence the total duration of the cycle is $\tau=\tau_u+\tau_b$. We consider a single ion in a harmonic trap as the the working substance of the system, while each bath consists of two lasers weakly coupled to the system, which raise and lower the occupation number of the quantum harmonic oscillator at different rates.  The overall effect of this weak system-bath interaction is such that after sufficient time has elapsed, the system becomes a thermal-like state at an effective temperature determined by a ratio involving the raising and lowering occupation number rates.

In section (\ref{sec:unitary}) we first analyze the limiting case of the ideal bath couplings where the system relaxes to final states that are effectively thermal-like at the end of the two dissipative strokes. 

In the following section (\ref{sec:dissipative}) we generalize our study to the case of an engine cycle in which the dissipative strokes are not coupled for long enough times to bring the system to a thermal-like state. Lastly,  we summarize our main findings and  present our conclusions in section (\ref{sec:conclusions}).

\section{Model}\label{sec:model}

We study as an idealized model for a quantum Otto cycle a one dimensional harmonic oscillator whose trapping frequency can be controlled in time and which is {\it weakly} coupled to external baths \cite{Alicki1979}, see Fig.~(\ref{fig:1}). We focus on a particular experimental realization made with a single ion in a Paul trap which can be  cooled and heated via side-band cooling by the use of two simultaneously acting lasers \cite{sidebandcooling}. The evolution of the density operator $\rhom(t)$ of the system to the external baths can thus be described by a master equation in (Markovian) Lindblad form \cite{GoriniSudarshan1976,Lindblad1976}, 
\begin{align}
 \frac{d\rhom}{dt} = -\frac{\im}{\hbar}\left[\Hop(t),\rhom\right] + \Dop(\rhom,t) \;.
 \label{eq:master}
\end{align}
The time-dependent system Hamiltonian is explicitly given by 
\begin{equation}
\Hop(t)=\left(\nop+\frac{1}{2}\right)\hbar\omega(t)\label{eq:Ham}
\end{equation}
Here, $\nop=\adop\aop$ is the number operator and $\aop$ ($\adop$) is the lowering (raising) operator, while $\omega(t)$ denotes the time-dependent frequency of the trap. The dissipator $\Dop(\rho,t)$ in (\ref{eq:master}) is given by \cite{CiracPhillips1992}:
\begin{multline}
\Dop(\rhom,t)  = \lambda^+(t) \left(2\adop\rhom\aop - \left\{\aop\adop,\rhom \right\}\right)
\\ + \lambda^-(t) \left(2\aop\rhom\adop - \left\{\adop\aop,\rhom \right\} \right) \;,
\label{eq:dissipator}
\end{multline}
where $\lambda^+$ and $\lambda^-$ denote the raising and lowering occupation rates respectively. Note that as a consequence of the protocol used for $\omega^2(t)$ (see Fig.~\ref{fig:1}), the system Hamiltonian does not change in time  during the dissipative parts of the cycle when the external baths are acting on the system. Given a fixed trapping frequency $\omega$, the dissipator tends to drive the system towards the diagonal quantum state that assumes the form
\begin{equation}
\rhom=\sum_n\rho_{nn} |n\ran \lan n|=\sum_n\frac{e^{-(n+1/2)\beta\hbar\omega}}{Z(\beta,\omega)} |n\ran \lan n| \;,
\end{equation}
where $\rho_{nn}$ denotes the normalized occupation probability in state  $|n\ran$, being the eigenstate of the Hamiltonian (\ref{eq:Ham}) and the effective inverse temperature $\beta$ of the steady state is such that $\lambda^+/\lambda^-=e^{-\beta\hbar\omega}$. $Z(\beta,\omega)$ is the partition function of a 1D harmonic oscillator at inverse temperature $\beta$ and trapping frequency $\omega$, and is given by
\begin{equation}
Z(\beta,\omega)=\frac{{e^{-\beta\hbar\omega/2}}}{1-e^{-\beta\hbar\omega}}.
\end{equation}
The relative strengths of $\hr$ and $\lambda^-$ determine the effective temperature for the system, provided that the steady state is reached. We would like to emphasize that although the baths considered can prepare a single ion in a thermal-like state, they do not constitute actual thermal baths. Instead, they impose a certain distribution of occupation of the energy levels that is independent of the energy difference between the levels. Given the non-thermal property of the baths we prefer to refer more precisely to the specific steady state of the (time independent) master equation as `thermal-like' and to $\beta$ as an `effective' inverse temperature.\\

Next we consider the operation of the quanutm Otto engine in greater detail, see Fig.~\ref{fig:1}. The cycle consists of two {\it unitary} strokes, each followed by a corresponding dissipative stroke in which the system is weakly coupled to the environment while the system parameters are held fixed (iso-parametric processes) \cite{ZhengPoletti2014, ZhengPoletti2015b}. We stress that the dissipative coupling is based on the weak coupling assumption, where the raising and lowering rates are much smaller than the internal electronic levels of the ion. In complete analogy to the classical Otto cycle, no heat is thus being exchanged during the unitary strokes $1 \to 2$ and $3 \to 4$, while no work is done during the iso-parametric processes $2 \to 3$ and from $4 \to 1$ of the cycle (see below for a more detailed explanation).

The evaluation of quantum work fluctuations \cite{TalknerHanggi2007,Talknerpre2008,HanggiTalkner2016}  requires nonselective, projective energy measurements on the {\it total} combined system  composed of the system, the baths and the mutual system-bath interactions \cite{HanggiTalkner2016,Watanabepre2014}. The determination of work in a stroke operation thus mandates (nonselective) energy measurements to be applied before and after each stroke of the Otto cycle.

Typically this presents a formidable  challenge, both for theory and even more so for experiments. This difficult task persists {\it even} in the case in which the coupling among the baths and the system of interest is weak \cite{Campisi2009}. Namely, with all energetic contributions of the system-bath interactions being neglected, the work fluctuations for the system  are still composed of both the changes in the internal energy of the system and, in general, finite energy exchanges with the baths. 

Both these contributions are typically non-vanishing. However, the difficult task becomes feasible if for instance, the average heat exchange is vanishing, yielding an average work exchange $\lan W \ran$ that is equal to the change of the energy of the system $\lan\Delta E\ran$, i.e.  $\lan W \ran = \lan \Delta E\ran$ \cite{footnote}. Likewise, when the work on the total system is vanishing, it implies that the change of the energy of the system alone is determined solely by the typically quite intricate heat exchange $Q$ among the  baths and the system, i.e. $\lan \Delta E \ran = \lan Q \ran$. In this case the projective measurement of the (nonselective)  bare system energy alone is sufficient \cite{footnote}. It thus demands that the quantum state given by the corresponding reduced density operator must be calculated. This in turn allows the overall average exchanges of either work or heat to be evaluated. \\

We next introduce a nonselective energy postmeasurement following each stroke of the cycle. This can be formalized by writing down  the corresponding (postmeasurement) density operators explicitly. The average energies are obtained in terms of projective energy measurements of the corresponding quantum state of the Otto engine. By introducing the projection operator $\Pi_n(t)=|n(t)\ran\lan n(t)|$, where $|n(t)\rangle$ is the instantaneous $n$-th energy eigenstate of the corresponding time-frozen system Hamiltonian at time $t$, the effect of the postmeasurement on the state at time $t$ is then given by a nonselective quantum state, and the reduced density operator $\rhom(t^+)$ that reads 
\begin{equation}
\rhom(t^+)=\Me\big(\rhom(t)\big) = \sum_n \Pi_n(t) \rhom(t) \Pi_n(t)\;.  \label{eq:measure}
\end{equation}
Since we are primarily interested in studying time-asymmetric protocols, we consider cycles in which a total  time span $\tau_u$ is spent on the unitary strokes and a total time $\tau_b$ on the dissipative strokes. We further parametrize the distribution of the time intervals within the unitary and dissipative strokes to allow for asymmetry in the driving protocol. For instance the time spent on the compression stroke $1 \to 2$, is given by $t_2-t_1=r_u\tau_u$ where $r_u$ is a real number between $0$ and $1$, while the time spent on the expansion stroke, $3 \to 4$, is given by $t_4-t_3=(1-r_u)\tau_u$. Similarly, for the dissipative strokes, coupling to the hot bath $2 \to 3$ is performed in time $t_3-t_2=r_b\tau_b$ and to the cold bath $4 \to 1$, in time $t_1+\tau-t_4=(1-r_b)\tau_b$.

For the unitary strokes of the cycle, we consider a protocol in which $\omega^2(t)$ varies linearly between $\omega_1$ and $\omega_2$, such that  $\omega^2(t)=\omega_1^2 + (\omega_2^2-\omega_1^2)(t-t_{1})/r_u\tau_u$
for the compression stroke and
$\omega^2(t) = \omega_2^2 + (\omega_1^2-\omega_2^2)(t-t_{3})/[(1-r_u)\tau_u]$
for the expansion stroke. Here we have used $t_{1}$ and $t_{3}$ respectively for the times at which the system is at stages $1$ and $3$ of the cycle. A plot of the time dependence of $\omega^2(t)$ is given in Fig.~\ref{fig:1}(b) (continuous blue line).

On the dissipative end of things, the coupling to the hot bath is turned on instantaneously at time $t_{2}^+$ from zero to the values $\la^\pm_{3}$ and back to zero at time $t_{3}$. Similarly, on the cold end, the bath couplings are again instantaneously switched from zero to  $\la^\pm_{1}$ for the time between $t_{4}^+$ and $t_{1}+\tau$. A depiction of the change of $\lambda^{\pm}$ is given in Fig.~\ref{fig:1}(b), where the green-dashed line represents $\lambda^-$ and the dot-dashed red line depicts $\lambda^+$. Note that $\lambda^- > \lambda^+$ and that their ratio is different in the two dissipative strokes, indicating that the system is driven towards  different Gibbs-like states.

In the following we use the notation $S_{a\to b}(\rhom)$ for the map corresponding to the stroke from $a$ to $b$ acting on the state $\rhom$. Explicitly, considering a postmeasurement density operator $\rhom(t^+_1)$ for the system, just after an energy measurement, the various strokes are given by
\begin{align}
\rhom(t^+_2)&=S_{1\to 2}\left(\rhom(t^+_1)\right) \nonumber \\
&= M_{t_2}\Big(K_{t_1,t_2}\rhom(t^+_1)\Big) \label{eq:rho12}
\end{align}
where $K_{t,t'}(\rhom(t))=\hat{U}_{t,t'}\rhom(t)\hat{U}^{\dagger}_{t,t'}$ with $\hat{U}_{t,t'}=\mathcal{T}\exp\lp -\im \int_t^{t'} \Hop(s)ds  \rp$ and $\mathcal{T}$ denotes time ordering.
It then follows that
\begin{align}
\rhom(t^+_3)&=S_{2\to 3}\left(\rhom(t^+_2)\right) \nonumber \\
&= M_{t_3}(\La_{t_2,t_3,\omega_2,\la^\pm_3}\rhom(t^+_2))  \label{eq:rho23}
\end{align}
where $\La_{t_2,t_3,\omega_2,\la^\pm_3}$ is a non-unitary map that evolves a density operator $\rhom (t)$ from $t^+_2$ to $t_3$ using the dissipative Lindblad master equation (\ref{eq:master}-\ref{eq:dissipator}) with $\omega(t)=\omega_2$ and $\la^\pm(t)=\la^{\pm}_3$, where the parameters are time independent.
The cycle closes upon applying the last two strokes; i.e.,
\begin{align}
\rhom(t^+_4)&=S_{3\to 4}\left( \rhom(t^+_3) \right) \nonumber \\
& = M_{t_4}(K_{t_3,t_4}\rhom(t^+_3))  \label{eq:rho34}
\end{align}
and back to the initial steady state, a fixed point of the cycle composed of four strokes, i.e.,
\begin{align}
\rhom(t^+_1+\tau)= \rhom(t^+_1)&= S_{4\to 1}\left( \rhom(t^+_4) \right) \nonumber\\
&= M_{t_1}(\La_{t_4,t_1+\tau,\omega_1,\la^\pm_1}\rhom(t^+_4))  \label{eq:rho41}
\end{align}

The steady state of the map is thus used to characterize the cycle given by the combination of the four strokes, which in our case is unique, and is equivalent to the diagonal fixed point, obeying
\begin{align}
\rhom(t^+_1)= S_{4\to 1}\!\Bigg(\!S_{3\to 4}\!\bigg(\!S_{2\to 3}\!\Big(\!S_{1\to 2}\!\big(\rhom(t^+_1)\!\big)\!\!\Big)\!\!\bigg)\!\!\Bigg).  \label{eq:fixedpoint}
\end{align}

Because the compression and expansion strokes are unitary the system is isolated from the baths; i.e., no heat  can be  exchanged. In presence of vanishing heat the mean work output determines the average work via the sole  difference of average energies of the system. This implies that the average work for the compression and expansion strokes are defined by the average of the two projectively measured energies of the isolated system \cite{TalknerHanggi2007,HanggiTalkner2016}; i.e.,
\begin{align}
\langle W_{1\to 2} \rangle &= \langle E_{2} \rangle - \langle E_{1 }\rangle \;, \label{eq:W12}
\\ \langle W_{3 \to 4} \rangle &= \langle E_{4} \rangle - \langle E_{3 }\rangle \;, \label{eq:W34}
\end{align}
where the average energies are
\begin{align}
\lan E_i \ran = \tr\lp \Hop(t_i)\rhom(t_i)  \rp. \label{eq:energy}
\end{align}
Clearly the average energy can also be computed using the post measurement density operator giving $\lan E_i \ran  = \tr\lp \Hop(t_i)\rhom(t^+_i) \rp$.

Keeping in mind that the baths are  weakly coupled to the system, heat exchange with any bath is solely given by the negative of the corresponding bath energy changes. Moreover, during the strokes $2\to 3$ and $4\to 1$ the system Hamiltonian does not change; put differently, {\it no} work is applied on the {\it total} system composed of system and baths including the weak mutual interactions (energy conservation). Therefore, the average work applied to the system  with the control parameter for ${\hat H}$ held constant is vanishing as well \cite{CampisiTalkner2011,Campisi2009}. The balance of energies exchanged thus implies that average heat exchange follows from a corresponding change in bare system energy alone, assuming here that the system Hamiltonian is not dressed by its interaction with the environment \cite{couplingprl}. The average system energies are  evaluated from the corresponding quantum state reached at the corresponding times $ \{t_i\}$. These mean values follow from the set of projective measurements of the system Hamiltonian by use of the corresponding reduced density operator for the system at time ${\hat{\rho}(t_i)}$.

Accordingly, we  hence find that the average values of heat exchanged are determined by 
\begin{align}
\langle Q_{2 \to 3} \rangle &=  \langle E_{3} \rangle - \langle E_{2 }\rangle \label{eq:heat23} \;, \\
\langle Q_{4 \to 1} \rangle &=  \langle E_{1} \rangle - \langle E_{4 }\rangle \;. \label{eq:heat23}
\end{align}
The net work of the cycle is thus given by
\begin{align}
\langle W \rangle =\langle W_{1\to 2} \rangle + \langle W_{3\to 4} \rangle \label{eq:network}
\end{align}

The efficiency of this cycle can be appropriately defined as the ratio involving the net average work output $\langle W \rangle$ divided by the net heat transferred from the hot bath into the system, $Q_{ 2\to 3}$. Note that here we adopt a negative sign convention for the net work, because we are primarily interested in a thermodynamic engine which does work on a load. The efficiency $\eta$ for this quantum Otto cycle is thus given by 
\begin{align}
\eta &=-\frac{\langle W \rangle}{\langle Q_{2 \to 3} \rangle} \;. \label{eq:efficiency}
\end{align}

\section{Optimal time distribution between unitary strokes} \label{sec:unitary}

Next, we numerically investigate different scenarios in which the operations are performed in finite-time. For details of our numerical analysis, we refer the readers to Appendix A. We begin by first considering the cases in which the time spans of the processes $2 \to 3$ and $4 \to 1$ (i.e. $r_b\tau_b$ and $(1-r_b )\tau_b$) are both sufficiently long, such that we can safely assume that the quantum states $1$ and $3$, after the respective dissipative strokes, are given by a thermal-like quantum state; i.e.,
\begin{equation}
\rhom_{i}=\sum_n{\frac{e^{-(n+1/2)\beta_{i} \hbar\omega_i}}{Z(\beta_i,\omega_j)} |n\rangle\langle n|}
\end{equation}
where $i=1,3$, $\omega_3=\omega_2$. Upon combining Eqs.~(\ref{eq:rho12}), (\ref{eq:rho34}) and (\ref{eq:W12}-\ref{eq:energy}), it is possible to compute $\lan W_{1\to 2}\ran$ and $\lan W_{3\to 4}\ran$. Note  \cite{DeffnerLutz2008, RezekKosloff2009} for analytical expressions of the mean work for various forms of $\omega(t)$.

%
%%%%%%%%%%%%%%%%%%%%%%%%%
% Fig2
%
\begin{figure}
\includegraphics[width=\columnwidth]{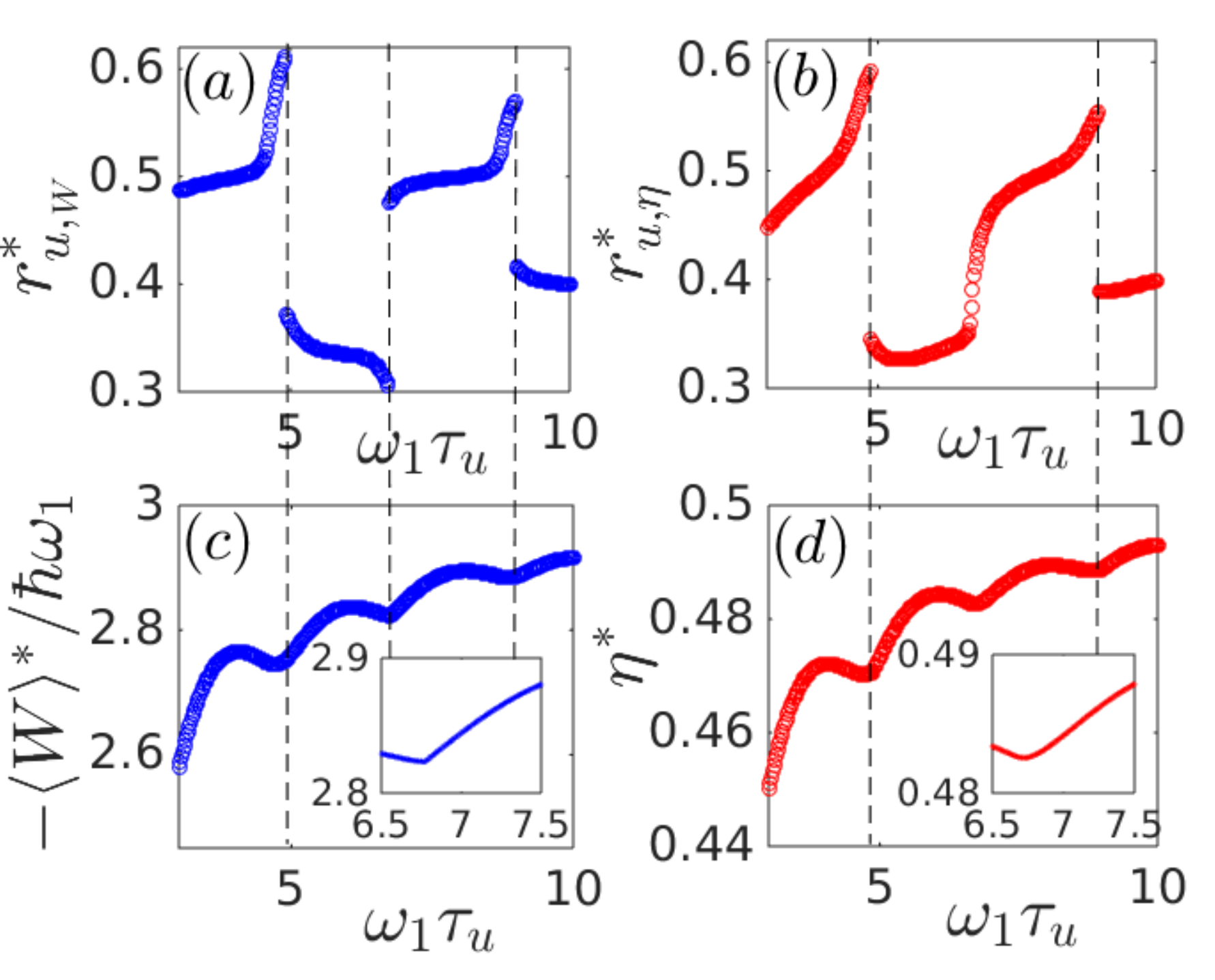}
 \caption{(color online) {\bf Optimal distribution of times between strokes.} Optimal values of the unitary parameter (a) $r^{*}_{u,_W}$ and (b) $r^*_{u,\eta}$ that optimize respectively, the work extracted $\langle W \rangle/\hbar \omega_1 $ and efficiency $\eta$ versus $\tau_u$. (c) The optimal work, $\lan W\ran^*$, corresponding to the values of $r^*_{u,_W}$ in (a) as a function of $\tau_u$. (d) Maximal efficiency $\eta^*$ corresponding to the values of $r^*_{u,b}$ in (b) a function of $\tau_u$. Hamiltonian and bath parameters are $\omega_2=2\omega_1, \beta_{1}\hbar\omega_1=0.5, \beta_{3}\hbar\omega_1=0.1$ for all cycles. The insets in (c,d) are close-ups of the respective quantities at $\omega_1\tau_u \approx 6.8$, where $r^*_{u,_W}$ changes discontinuously while $r^*_{u,\eta}$ changes smoothly. We note that the derivative of the optimal work $\lan W \ran^*$ changes abruptly while that of the maximum efficiency $\eta^*$ is smooth. The vertical dashed black lines indicate the position of the jumps of $r^*_{u,_W}$ and $r^*_{u,\eta}$.
}
\label{fig:2}
\end{figure}

\begin{figure}
\includegraphics[width=\columnwidth]{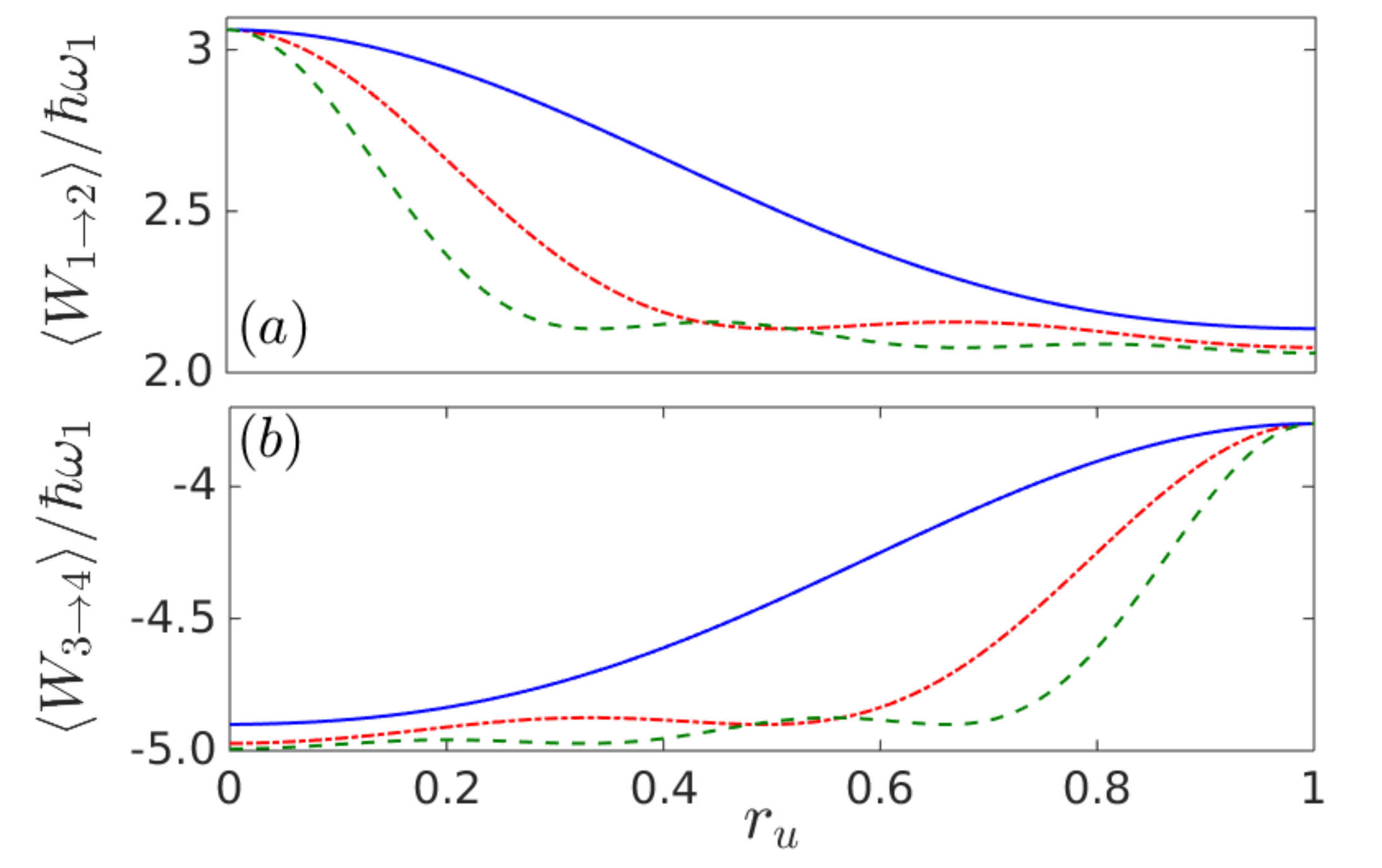}
\caption{(color online) {\bf Work transfer of a single compression and expansion process.} (a) $\langle W_{1\to 2} \rangle / \hbar\omega_1$ and (b) $\langle W_{3 \to 4} \rangle / \hbar\omega_1$  as a function of $r_u$ for the compression arm of an Otto cycle that operates between $\omega_1$ and $\omega_2=2\omega_1$ with $\tau_u\omega_1=2$ (continuous blue line), $\tau_u\omega_1=4$ (dash-dotted red), $\tau_u\omega_1=6$ (dashed green). Inverse temperatures of the bath are held at $\beta_{ 1}\hbar\omega_1=0.5$ and $\beta_{3 }\hbar\omega_1=0.1$ respectively for the compression and for the expansion processes. }
\label{fig:3}
\end{figure}

We next search for the optimal distribution of the total time spent on the unitary processes $\tau_u$ between the two strokes, which is parametrized by $r_u$. For instance, for $r_u=0$ the stroke from $1 \to 2$ constitutes an abrupt quench while a total of time $\tau_u$ is spent in the stroke $3 \to 4$. For larger values of $r_u$ the time spent in the stroke from $1 \to 2$ increases while the one from $3 \to 4$ decreases until $r_u=1$, i.e.,  when this last stroke becomes an abrupt quench. The ratio $r_u$ for which the net work is optimum is referred to as  $r^*_{u,_W}$ and we denote the maximum work as $\lan W \ran^*$, given by
\begin{align}
-\lan W\ran^* = \max_{r_u}\left(-\lan W\ran\right) \label{eq:maxwork}
\end{align}
(we remind the reader that the net work for an engine is negative). We depict with Fig. \ref{fig:2}(a) the value of $r^*_{u,_W}$ as a function of the total time spent on the unitary strokes $\tau_u$. We observe that at some critical values of $\tau_u$, discontinuities in the ratio $r^*_{u,_W}$ occur. At the occurrence of these jumps, also the derivative of the net work output changes abruptly, see Fig.~\ref{fig:2}(c) and its inset.

A similar result emerges when $r_u$ is chosen so as to maximize the efficiency, denoted here as $r^*_{u,\eta}$, see Fig.~\ref{fig:2}(b). The corresponding optimal efficiency is denoted as $\eta^*$ and it is 
\begin{align}
\eta^* = \max_{r_u}\left(\eta\right). \label{eq:maxeffi}
\end{align}
The maximum efficiency $\eta^*$ as a function of $\tau_u$ is shown in Fig.~\ref{fig:2}(d). In Fig.~\ref{fig:2}(b) we observe that the abrupt jumps are also present in the values of $r^*_{u,\eta}$, although they occur at different values of $\tau_u$ from the jumps in $r^*_{u,W}$. Moreover, in certain regions of $\tau_u$ an abrupt jump of $r^*_{u,_W}$ can occur while a continuous variation of $r^*_{u,\eta}$ occurs, as is the case for $\omega_1\tau_u\approx 6.8$, see the insets of Fig.~\ref{fig:2}(c,d). Hence, given a total time $\tau_u$ for the unitary strokes, the optimal distribution of times between the strokes $1\to 2$, e.g. $r_u\tau_u$, and $3 \to 4$, e.g. $(1-r_u)\tau_u$, depends on the quantity being optimized for; For instance either the net output work or the efficiency. Also note that values of our numerically evaluated optimal efficiency are always below that of the Carnot bound $\eta_C=1-\frac{\beta_3}{\beta_1}=0.8$ as they should be.

The nonlinear behavior in the optimal ratio $r^*_{u,_W}$ and $r^*_{u,\eta}$ can be understood by analyzing the work output from a single compression $\langle W_{1\to 2} \rangle$, or expansion process $\langle W_{3 \to 4} \rangle$. As can be seen from Fig.~\ref{fig:3} (see also \cite{AcconciaDeffner2015}), the work transferred in a single unitary stroke becomes a non monotonic function of the ratio $r_u$ when the value of $\tau_u$ becomes sufficiently large. This behavior is ultimately responsible for the phenomenon we observe. In fact, finding the optimal time spans between the strokes stems from matching the optimal work output from two non-monotonic functions of time under the constraint of a given total time spent on the two unitary strokes. In Fig.~\ref{fig:3} we observe that the number of oscillations present in the work output as a function of $r_u$ increases with the  duration of total time spent on the two unitaries $\tau_u$

%%%%%%%%%%%%%%%%%%%%%%%%%
% Fig4

\begin{figure}
\includegraphics[width=\columnwidth]{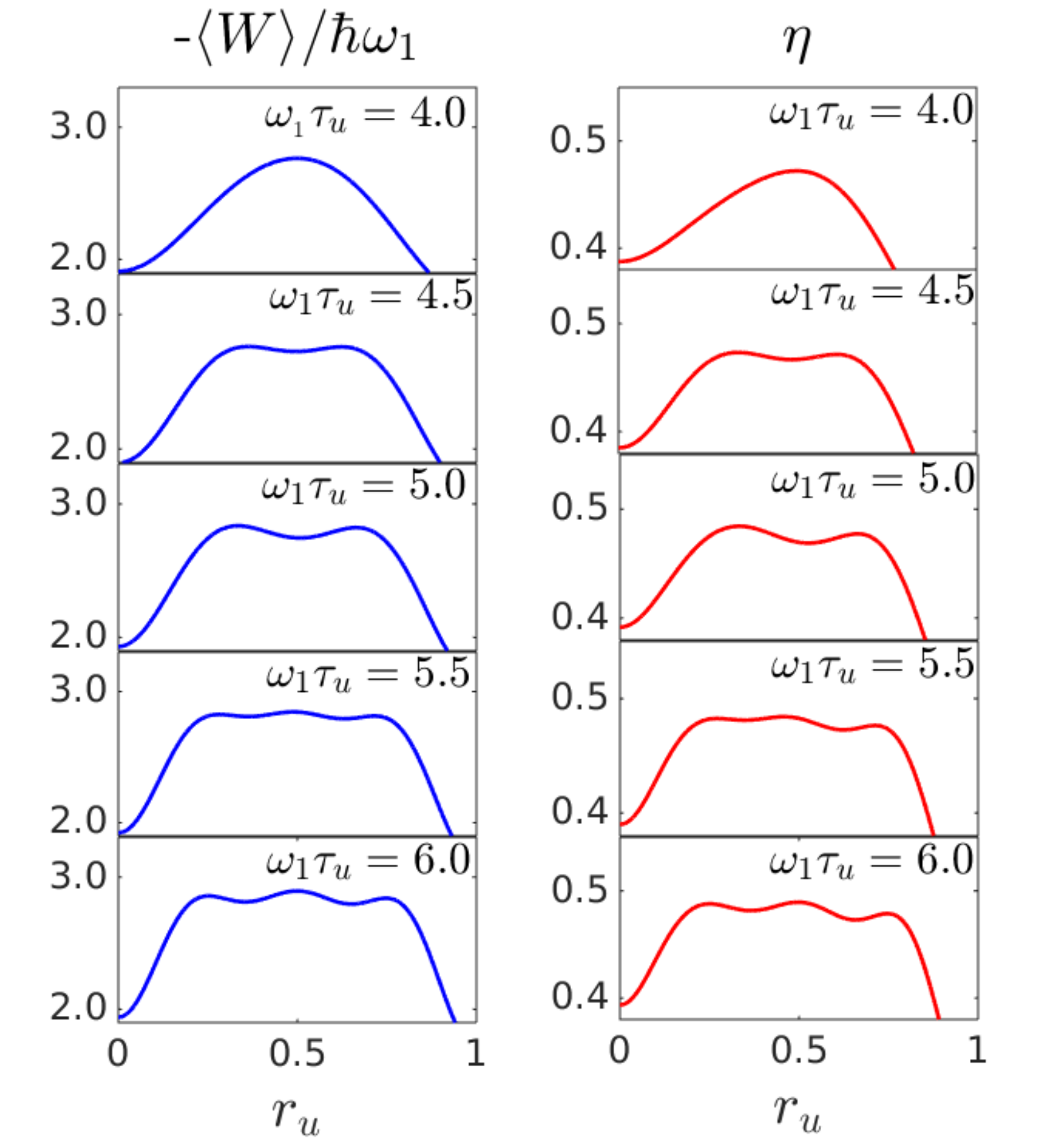}
 \caption{(color online) {\bf Oscillations of the net work and efficiency.} (Left panels) Mean work extracted and (Right panels) efficiency of cycles as a function of $r_u$ for cycles with the parameters set at $\omega_2=2\omega_1, \beta_{1}\hbar\omega_1=0.5, \beta_{3}\hbar\omega_1=0.1$. Increasing number of oscillations in both net work and efficiency with the unitary timescale $\tau_u$ results in discontinuities in $r_{u,_W}^{*}$ and $r^*_{u,\eta}$.} \label{fig:4} 
\end{figure}

In Fig.~\ref{fig:4}, the value of average work (left) and efficiency (right) as a function of the ratio $r_u$ are shown for different values of $\tau_u$. Since the work in each stroke is a non-monotonic function of the time spent, the net work, which is the sum of work in the two unitary strokes becomes an oscillating function. 
For larger $\tau_u$-values  the number of local minima or maxima of the net work output, or of the efficiency, increases as the total time in the unitary strokes increases. The emergence of a new global extremum can occur either via the increase in magnitude of a local extremum, such that it becomes the global one, or when the global extremum turns unstable and in turn yields two extrema, with one of the two becoming the new global extremum. The former route is reminiscent of the behavior of the free energy as a function of the order parameter as temperature changes in an ordinary first order phase transition, while the latter mimics the behavior of a second order phase transition.

%%%%%%%%%%%%%%%%%%%%%%%%%
% Fig5
\begin{figure}
\includegraphics[width=\columnwidth]{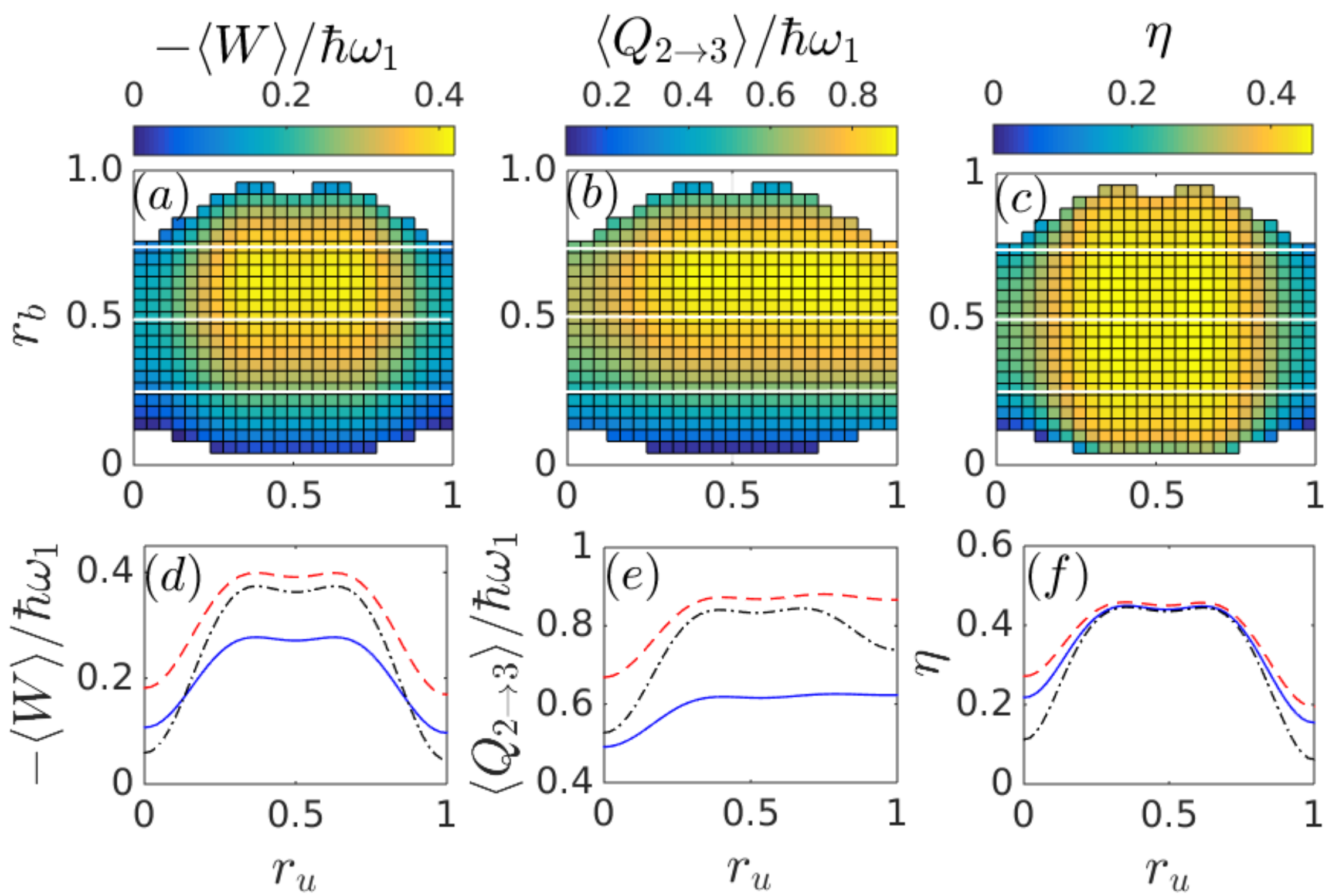}
 \caption{(color online) {\bf Cycle Performance at short bath coupling timescales.} (a) Average net work $\langle W \rangle$, (b) average heat from the hot bath $\langle Q_{2 \to 3} \rangle$ and (c) efficiency $ \eta $ of Otto cycles where $\omega_1\tau_u=\omega_1\tau_b/10=5$ as a function of $r_u$ and $r_b$. Regions that are shaded white are domains in which the cycle is non physical as it is not doing work but instead receiving it. Plot (d-f) show the horizontal cuts highlighted by white lines in figures (a-c). In particular, in (d-f) the blue continuous line corresponds to $r_b=0.25$, the red dashed line to $r_b=0.5$, and the black dot-dashed line to $r_b=0.75$. We use $\lambda^-_{1,3}=\omega_1/10$, $\lambda^+_3=\lambda^-_3/e$ and $\lambda^+_1=\lambda^-_1/e^{10}$ which corresponds to $\beta_{1}\hbar\omega_1=10$ and $\beta_{3}\hbar\omega_1=0.5$.  } \label{fig:5}
\end{figure}

\section{Optimal time distribution within unitary and dissipative strokes} \label{sec:dissipative}

In the previous section, the state at the beginning of the compression and expansion strokes, $1$ and $3$ respectively, were assumed to be effectively thermal-like (because enough operation time was spent on the two dissipative strokes $2 \to 3$ and $4 \to 1$). We now consider the case in which the time spans of the dissipative strokes, $r_b\tau_b$ and $(1-r_b)\tau_b$, are too short for thermalization to occur such that the quantum states in $1$ and $3$ are no longer thermal-like. Under such circumstances, as detailed with Eq. (\ref{eq:fixedpoint}), $\rhom(t_i^+)$ is the (unique) fixed point of the four strokes which connects the quantum state $1$ back to itself.

Since after each energy measurement the density operator is diagonal in the basis of instantaneous energy eigenstates, computing $\rhom(t_1^+)$ amounts to finding the eigenvector associated to the eigenvalue $1$ of the corresponding Markovian map in Eq. (\ref{eq:fixedpoint}). In particular, we would like to stress that the numerical evolution of Eq. (\ref{eq:master}) in the strokes $2 \to 3$ and $4 \to 1$ is particularly simple because after the measurements in $2$ and $4$ the density operator is diagonal in the instantaneous eigenbasis of $\Hop(t_i^+)$. Moreover, our  specific choice of dissipator $\mathcal{D}$, given by Eq.(\ref{eq:dissipator}), preserves the diagonal form when acting on a diagonal density operator. This implies that the density operator, in the strokes $2\to 3$ and $4 \to 1$ commutes with the Hamiltonian and hence its evolution only depends on the part involving the dissipation.

In order to stay in the weak coupling regime throughout the evolution, we ensure  that the ratio $\lambda^{\pm}_i / \omega_1$ is always much lesser than $1$. However, given the stylized setup of our system, keeping the products $\lambda_3^{\pm}r_b\tau_b$ and $\lambda_1^{\pm}(1-r_b)\tau_b$ fixed while varying $\lambda_i^{\pm}$, $r_b$ and $\tau_b$ individually, results in the same $\hat{\rho}(t_i^{+})$ obtained and hence leads to the same net work output and efficiency for the cycle \cite{validity}.

In Fig.~\ref{fig:5} we depict the color coded intensity  for net average work $\lan W\ran$, average heat input $\lan Q_{2 \to 3}\ran$ and the efficiency $\eta$ [cf. in Fig.~\ref{fig:5}, panels $(a)$, $(b)$ and $(c)$]. We choose $\tau_u=\tau_b/10=5/\omega_1$ because this is a long enough time to obtain two extrema in the cycle work output when the total time is distributed across the strokes (there would be only one maximum for shorter timescales). It should also be noted that in Fig.~\ref{fig:5} some regions are colored in white. These  regions correspond to cases (similarly to \cite{FeldmannKosloff2003, FeldmannKosloff2012, FeldmannKosloff2015}) for which the finite-time operation of the cycle  does not yield an overall  negative net work output. Notably, these regions occur when any one of the dissipative or unitary strokes is performed in too short a time. In Fig.~\ref{fig:5} we observe a quantitative change of the mean work output, heat and efficiency due to the finite time spent on the dissipative strokes, but do not find any qualitative changes.
This is highlighted in Fig.~\ref{fig:5}(d-f) by the horizontal cuts of the intensity maps in Fig.~\ref{fig:5}(a-c) for three different values of $r_b$.

\section{Conclusions}\label{sec:conclusions}

With this work we investigated the efficiency, net work output and input heat for a quantum Otto engine operated in a finite time. The cycle we consider is composed of two unitary strokes connected by two dissipative strokes. During the total time $\tau=\tau_u + \tau_b$ of the cycle operation, a typically asymmetric portion is spent on the two unitary strokes of total duration $\tau_u $ and the remaining time span $\tau_b$ on the two dissipative strokes. We numerically evaluated the optimal distribution of time spans within the unitary and dissipative strokes while optimizing either the net work output or the efficiency. The distribution of times is parametrized by $r_u$ and $r_b$ such that the time in the stroke $1 \to 2$ is $r_u\tau_u$ and $2 \to 3$ is $r_b\tau_b$. In Sec.~\ref{sec:unitary} we elaborated on the case in which the time spent in the dissipative strokes is long enough such that the baths effectively thermalize the system at the end of the stroke. 

We observed that optimizing for the work output results in discontinuous jumps in $r^*_u$ across values of $\tau_u$. Likewise, optimizing for the efficiency results in a similar jump behavior, albeit in a different location along $\tau_u$. This feature stems from a non-monotonic dependence of work output on the time spent in each unitary  stroke. This phenomenon  is  present as well when the baths do not fully thermalize the system as shown in Sec.~\ref{sec:dissipative}.   

We would like to note that the engine cycle considered includes energy measurements at the end of each stroke. Because in quantum mechanics each measurement affects the system via a back action, and since energy measurements are necessary to obtain the net energy balance for work output and heat exchange, it is essential to detail these measurements after each individual stroke of the cycle. We also stress that in any of these strokes we encounter non-equilibrium scenarios as the energy balance is {\it not} between corresponding thermal equilibrium states. Put differently, all our energy balance relations are manifestly non-equilibrium  relations that cannot be labelled `thermodynamic first law' relations.  The latter involves knowing the difference between two internal energy state functions.

In contrast to the the quasi-static and reversible Otto cycle in thermal equilibrium such features of abrupt jumps are absent. Therefore, it would be of interest to see if the features as depicted in our set of figures, are indeed present in an experiment of a quantum Otto cycle.   

The study of finite-time quantum engine cycles is necessary for the implementation of such systems. 
As exemplified by this work, the dynamics involved in finite time quantum engine cycles is rich and should be studied thoroughly.   \\   

{\bf Acknowledgments} D.P. acknowledges support from Singapore MOE Academic Research Fund Tier-2 project (Project No. MOE2014-T2-2-119, with WBS No. R-144-000-350-112) and fundings from SUTD-MIT IDC (Project No. IDG21500104). Y.Z. is supported by NTU SUG M4081346. D.P. also acknowledges fruitful discussions with F. Binder, S. Fazio, J. Goold, K. Modi, S. Vinjanampathy.

\setcounter{equation}{0}
\renewcommand{\theequation}{A.\arabic{equation}} 
\setcounter{figure}{0} 
\renewcommand{\thefigure}{A.\arabic{figure}}

\begin{appendix}

\section{Numerical computations}

After each stroke, inclusive of the energy measurement, the density operator $\rhom$ is diagonal in the instantaneous energy eigenbasis and can thus be written as 
\begin{align} 
\rhom(t_i^+) = \sum_n \rd_n(t_i^+) \Pi_n(t).  
\end{align} 
where $\rd_n$ is the $n$-th element of the diagonal density operator. For a unitary stroke from $t_i^+$ to $t_{i+1}^+$ we get, using Eq.(\ref{eq:rho12}) or (\ref{eq:rho34}) 
\begin{align} 
\rd_n(t_{i+1}^+) = \sum_m \Pm^{n,m}_{t_i,t_{i+1}}\;\rd_m(t_i^+)   
\end{align} 
where 
\begin{align}
\Pm^{n,m}_{t_i,t_{i+1}} = |\lan n(t_{i+1})|\hat{U}_{t_i,t_{i+1}}|m(t_i)\ran|^2. 
\end{align}
The dissipative evolution for $t\in [t_i^+,\;t_k^+)$, due to Eqs.(\ref{eq:master}-\ref{eq:dissipator}) is instead given simply by   
%\begin{align} 
%\rd_n(t_{i+1}^+) = \sum_m \Dm^n_m\;\rd_m(t_i^+)   
%\end{align} 
\begin{align} 
\frac{d\rd_n(t)}{dt} = 2& \left\{ n \lambda_{k}^+  \rd_{n-1}(t) + (n+1) \lambda_{k}^-  \rd_{n+1}(t)   \right.\nonumber\\ 
&-\left.\left[   (n+1)\lambda_k^+ +n\lambda_k^-       \right]\rd_n(t)\right\} 
\end{align} 
where $i=2$ or $4$ and $k=\left[(i+1)\mod(4)\right]$. This simple form of time evolution is due to the fact that the density operator $\rhom$ is diagonal in the instantaneous energy eigenbasis at all instances of the dissipative strokes and hence it commutes with $\Hop$. In our implementation we have kept $n=50$ eigenstates of the harmonic oscillator, which is sufficient for the effective temperatures and dynamics involved. 

In our numerical simulations, we have rewritten Eq.~(\ref{eq:master}-\ref{eq:dissipator}) in terms of dimensionless parameters. Explicitly, by dividing Eq.~(\ref{eq:master}) by $\omega_1$, we obtain 
\begin{align}
 \frac{d\rhom}{d\tilde{t}} = -\im\left[\tilde{H}(\tilde{t}),\rhom\right] + \tilde{\Dop}(\rhom,\tilde{t}) \;.
\end{align}
where $\tilde{t}=\omega_1 t$, 
\begin{align} 
\tilde{H}=\Hop/\hbar\omega_1 = (\nop+1/2)\tilde{\omega}, 
\end{align}  
and  
\begin{multline}
\Dop(\rhom,\tilde{t})  = \tilde{\lambda}^+(\tilde{t}) \left(2\adop\rhom\aop - \left\{\aop\adop,\rhom \right\}\right)
\\ + \tilde{\lambda}^-(\tilde{t}) \left(2\aop\rhom\adop - \left\{\adop\aop,\rhom \right\} \right) \;, \nonumber 
\label{eq:dissipator_rescaled}
\end{multline}
where $\tilde{\omega}=\omega/\omega_1$ and $\tilde{\lambda}^{\pm}=\lambda^{\pm}/\omega_1$. As a consequence the inverse temperature $\beta$ can be written in dimensionless form $\tilde{\beta}=\beta\hbar\omega_1$ because 
\begin{align} 
\beta\hbar\omega=\left(\beta\hbar\omega_1\right)\left(\omega/\omega_1\right)=\tilde{\beta}\tilde{\omega}.
\end{align}

\end{appendix}


\begin{thebibliography}{99}
\bibitem{validity} It should be stressed again that the master equation (\ref{eq:master}), and the computation of heat and work in the dissipative strokes, are valid only for small $\lambda_i^{\pm}$.
\end{thebibliography}
\end{document}